\documentclass[prl,aps,showpacs,twocolumn]{revtex4}
\usepackage{epsfig}
\usepackage{latexsym}
\usepackage{graphicx}
\usepackage{amssymb}

\begin{document}
\title{Construction and Refinement of Coarse-Grained Models}

\author{Xin Zhou 
}

\affiliation{
Asia Pacific Center for Theoretical Physics
and Department of Physics, Pohang University of Science and Technology,
Pohang, Gyeongbuk 790-784, Korea
}


\date{\today}

\begin{abstract}
A general scheme, which includes constructions of coarse-grained (CG) models, weighted ensemble dynamics (WED) simulations and cluster analyses (CA) of stable states, is presented to detect dynamical and thermodynamical properties in complex systems. 
In the scheme, CG models are efficiently and accurately optimized based on a directed distance from original to CG systems, which is estimated from ensemble means of lots of independent observable in two systems. 
Furthermore, WED independently generates multiple short molecular dynamics trajectories in original systems. 
The initial conformations of the trajectories are constructed from equilibrium conformations in CG models, and the weights of the trajectories can be estimated from the trajectories themselves in generating complete equilibrium samples in the original systems. 
CA calculates the directed distances among the trajectories and groups their initial conformations into some clusters, which correspond to stable states in the original systems, so that transition dynamics can be detected without requiring a priori knowledge of the states. 
\end{abstract}


\maketitle 

Atomistic molecular dynamics (MD) simulations are very powerful tools to accurately evaluate dynamics and thermodynamics properties of complex systems, however they are limited to systems with small size and phenomena of short time. Recently, many different multiscale techniques are developed to extend the temporal/spatial scales of simulations
 ~\cite{Elber2005,VoterMG2002,BolhuisCDG2002,SherwoodBS2008}.   
Coarse-grained (CG) modeling, which reduces degrees of freedom and parameterizes effective interactions, offers a promising way to surmount the limitations 
~\cite{SherwoodBS2008,MuellerPlathe2002,KremerGroup,IzvekovV2005,ZhouJZR2008}. 
 The effective interaction of CG models, such as $U(x)$, is usually required to match the free energy surface, 
 $F(x) = -\ln \int e^{-V(r)} \delta(x-x(r))dr$,
in whole the CG conformational space, $x$. Here $x=x(r)$ are conformational functions in the original system with potential energy surface, $V(r)$. $\delta(\cdots)$ is the Dirac-$\delta$ function, and   
 $k_{B} T$ is set as the unit of energy. 
In CG approaches, some assumptions and approximations are inevitably introduced, because it is very difficult (if not impossible) to get an analyzed $F(x)$ in the high-dimension space, $x$. In the other hand, some interesting properties, such as transition dynamics, may be changed in the CG approaching. It is important to high efficiently construct CG models and to refine the CG models  
while it is necessary. 

A key of CG approaches is to define a cheap and accurate distance between CG models, such as $U(x)$, and original systems, such as $V(r)$,  or the corresponding free energies, $F(x)$. 
In traditional CG approaches~\cite{MuellerPlathe2002}, the distance 
is defined from ensemble means of some (arbitrarily) selected variables in the two systems. For example, ones calculate the difference of radial distribution function $g(z)$ in $U(x)$ and $V(r)$ and define the distance as $D_{trad}= \int dz \rho(z) [\langle g(z)\rangle_{U} - \langle g(z)\rangle_{V}]^{2}$~\cite{MuellerPlathe2002,KremerGroup}. Here $\rho(z)$ is an optional weight, and $\langle \cdots \rangle_{\gamma}$ are   ensemble means.
 In more recent works~\cite{IzvekovV2005,ZhouJZR2008}, ones directly estimate values of $F(x)$ or its gradients $\frac{\partial F}{ \partial x}$ at some sampled CG conformations, such as $x^{i}, i=1, \cdots, M$, then define the distance as the mean of $\delta V(x) = F(x)-U(x)$ or its derivative in the sample, {\it i.e.}, $D_{FE} = \frac{1}{M} \sum [\delta V(x^{i})]^{2}$ or $D_{FED}= \frac{1}{M} \sum [\frac{\partial \delta V(x)}{\partial x}]^{2}_{x^{i}}$. 
While $D_{FE}$ or $D_{FED}$ takes into account the overall characteristic of $F(x)$, the calculation of $F(x^{i})$ or $\frac{\partial F}{\partial x^{i}}$ is usually very time-consuming. 
In contrast, the traditional CG approaches are easier to be calculated but effects due to the arbitrary selection of variables are not very clear.  
 
In this letter, we first define a directed distance between any two systems based on ensemble means of a complete basis function set, then   
estimate the distance by calculating the means of some interesting observable in large conformational samples of the two systems. 
Thus we can efficiently and accurately optimize parameters of effective interactions of CG models by minimizing the directed distance. 
Furthermore, we present weighted ensemble dynamics (WED) simulations and cluster analyses to refine CG models to exactly reproduce dynamics and thermodynamics in original systems. WED randomly CG conformations and arbitrarily adds into the missing degrees of freedom with short relaxations to form initial conformations, then independently generates multiple short molecular dynamics simulations in the original systems. 
Besides statistically detecting ensemble dynamics in the original systems, WED reproduces the equilibrium properties by weighting these trajectories. The weights, which are independent on the short relaxation simulations,  
can be estimated from the trajectories themselves, or from a self-consistent equation based on cluster analyses (CA) of the trajectories. Without requiring a priori knowledge of stable states, we calculate directed distances among the trajectories and group their initial conformations into clusters ({\it i.e.} stable states), and identify transition trajectories among the stable states to detect the corresponding transition dynamics.  
The CG-WED-CA scheme provides a complete way in analyzing stable states, detecting transition dynamics, as well as enhanced sampling in complex systems. 
 
A natural definition of the distance between two potentials, such as $U(x)$ and $F(x)$, may be their overlap, $d(U,F)=\langle \phi_{U}(x)| \phi_{F}(x) \rangle$. Here $\phi_{F}(x) \propto e^{-F(x)/2}$ and $\phi_{U}(x) \propto e^{-U(x)/2}$ have already been normalized. 
However, it is difficult to use the overlap to parameterize effective potentials of CG models. 
Alternately, we define a directed distance 
$s^{2}_{F,U} \equiv \langle [\delta_{F} {\cal W}_{F,U}(x)]^{2} \rangle_{F}$,  
where the weight function ${\cal W}_{F,U}(x) \propto e^{F(x)-U(x)}$, $\delta_{F} A(x) \equiv A(x) - \langle A(x) \rangle_{F}$, 
and $\langle {\cal W}_{F,U}(x) \rangle_{F}=1$ without losing any generality. 
For any variable $A(x)$, 
\begin{eqnarray}
|\langle A(x) \rangle_{U} - \langle A(x) \rangle_{F} | \leq  \sigma ~ s_{F,U},
\label{error}
\end{eqnarray} 
where $\sigma$ is the fluctuation of $A(x)$ in the $F$ system. 
Thus $s_{F,U}$ measures the deviation of $U(x)$ from $F(x)$, since it provides a upper limit of errors in calculating ensemble means of any thermodynamical variable. 
While $\delta_{F}{\cal W}_{F,U}(x) \ll 1$, the directed distance is equivalent to the overlap, $d(U,F)$. 

We expand ${\cal W}_{F,U}(x)$ in an arbitrary complete basis set, such as, $\{A^{\mu}(x)\}$, 
\begin{eqnarray}
{\cal W}_{F,U}(x) =1 + g_{\mu\nu}(F)~\langle \delta_{F}A^{\mu}(x) \rangle_{U}~\delta_{F}A^{\nu}(x), 
\label{omega-expansion}
\end{eqnarray}
where 
$g_{\mu\nu}(F)$ is the inverse matrix of the variance-covariance matrix of basis functions, 
$g^{\mu\nu}(F) \equiv \langle \delta_{F}A^{\mu}(x)~\delta_{F}A^{\nu}(x) \rangle_{F}$. 
Here we used the Einstein summation notation. 
Thus, 
\begin{eqnarray}
s^{2}_{F,U} = g_{\mu\nu} a^{\mu} a^{\nu},  
\label{s2distance}
\end{eqnarray}
where $g_{\mu\nu}$ and $a^{\mu}=\langle A^{\mu} \rangle_{U} - \langle A^{\mu} \rangle_{F}$ are simple notations of $g_{\mu\nu}(F)$ and $\langle \delta_{F}A^{\mu}(x) \rangle_{U}$, respectively. 
In this letter, we generally use $U(x)$ to denote effective interactions of CG models, $V(r)$ to original systems, and $F(x)$ to free energy surfaces of $V(r)$ in the $x$ space. We also denote the missing degrees of freedom in the CG approaches as $y$, so that $(x,y)=r$. 
In principle, eq.(\ref{omega-expansion}) is exact and independent on the applied $U(x)$. An analyzed 
free energy $F(x)$ in any high-dimension space can be obtained by calculating the ensemble means of basis functions in $V(r)$ and in an arbitrary $U(x)$. 
In practice, because the ensemble means are estimated in finite-size samples, and a finite subset of the basis set is applied instead of whole the basis set, only an approximated $F(x)$, 
is obtained in the expansion. 
However, the directed distance $s_{F,U}$ provides a good approximation of the deviation of $U(x)$ from $V(r)$, if most interesting observable are included in the basis set.
Based on the directed distance, 
CG models are constructed as below: (1) sample $M$ conformations in $V(r)$ (or in a reference system if the simulations in $V(r)$ is too expensive), and calculate the means and variance-covariance matrix of chosen basis functions in the conformational sample. 
Here each basis function corresponds to a $M-$dimension vector, thus the number of independent basis functions is not more than $M$. The applied basis vectors can be orthogonalized and normalized to make $g^{\mu\nu}$ be the unit matrix; 
(2) set initial values of parameters of $U(x)$;
(3) generate CG conformations in the current $U(x)$ and calculate $s_{F,U}$ (and its derivative to the parameters of $U(x)$, if it is needed);  
(4) optimize the parameters of $U(x)$ by minimizing $s_{F,U}$ in some standard techniques, {\it e.g.}, the conjugate gradient method.
In eq.(\ref{s2distance}), interesting and important observable should be included in the basis set, so that the formed CG model at least reproduce the observable very well. 
In comparison with $D_{trad}$ in the traditional CG approach~\cite{MuellerPlathe2002,KremerGroup} and $D_{FE}$ in the free-energy-based CG methods~\cite{ZhouJZR2008}, $s_{F,U}$ takes into account the pair correlation among the selected basis functions to capture overall characteristics of $F(x)$. 
 
We further refine the formed CG models by developing ensemble dynamics techniques~\cite{Voter1998,ShirtsP2001}. 
Instead of the normal long single-trajectory simulations, ensemble dynamics simulations generate independently multiple short molecular dynamics trajectories in distributing computers and statistically analyze dynamical behaviors of systems. For example, Pande {\it et al.} generated hundreds of thousand nanosecond-scale trajectories and found a few microsecond-scale folding events~\cite{foldingathome} in all-atomic protein models with explicit water molecules. However, the ensemble dynamics usually arbitrarily selects initial conformations of the simulations in known states ({\it e.g.}, the folded and unfolded states of proteins) to identify particular transitions within the total simulation time scale, the distribution of the conformations collected from all the trajectories may be unknown. 
 We present weighted ensemble dynamics (WED) simulations which independently generates trajectories same as the normal ensemble dynamics, but the initial conformations of the trajectories are constructed from equilibrium conformations in $U(x)$, by arbitrarily adding the missing degrees of freedom, $y$, with short relaxation simulations in $V(r)$. The trajectories contribute to equilibrium properties of $V(r)$, with weights, $\{w_{i}\}$, where  
\begin{eqnarray}
w_{i}^{-1} \sim  \frac{1}{ t - \Delta_{i} } \int_{0}^{t-\Delta_{i}}  {\cal W}_{V,U}(r_{i}(\tau)) d \tau.  
\label{trajectoryweight}
\end{eqnarray}
Here ${\cal W}_{V,U}(r) \propto e^{V(r)-U(x(r))}$, and $t$ is the length of trajectories. 
 $\Delta_{i} \ge 0$ is selected so that the obtained $w_{i}$ almost does not changes as varying $\Delta_{i}$. 
 
Let us verify WED by considering the ensemble of all MD trajectories with length $t$. 
Since the trajectories are same generated from the Newtonian (or Langevin) equation of motion, 
we have, (1) conformations in a trajectory have the same weight in contributing to equilibrium properties; (2) trajectories started from an initial conformation have the same weight in the contribution, if the initial velocities are unbiasedly formed from the Maxwell velocity distribution.  
We denote the sub-ensemble of the trajectories which started from an initial conformation, such as $r_{0}$, as $[r_{0};t]$, and denote the mean of any $A(r)$ in the sub-ensemble as $\langle A \rangle_{[r_{0};t]}$. 
While $t$ is not very short, the conformational distributions of sub-ensembles of the trajectories, which started from neighboring conformations, such as $[r_{0};t]$ and $[r_{0}^{\prime};t]$, may be identical, ({\it i.e.}, $\langle A \rangle_{[r_{0};t]} = \langle A \rangle_{[r_{0}^{\prime};t]}$ for any $A(r)$). 
In other words, these initial conformations are equivalent in the $t-$length MD simulations, they belong to the same stable conformational  region, wherein the MD simulations easily reach equilibrium within $t$.  
Here the conformational distribution in the sub-ensemble $[r_{0};t]$, $P_{[r_{0};t]}(r) = \frac{1}{t} \int G(r_{0},0;r,t^{\prime}) dt^{\prime}$, and $G(r_{0},0;r,t)$ is the propagator of the applied MD algorithm ({\it i.e.}, simulator).  
Therefore, we strictly define stable conformational regions in any time $t$, without requiring to analyze if the time scales of the simulation dynamics are separated well, or if some particular trajectories already happened transitions. 
We conclude that all the trajectories started from a stable conformational region have the same weight in contributing to equilibrium properties, but trajectories started from different stable regions might have different weights. 
The total weights of the trajectories from a stable region should be proportional to the free energy of the stable region.
It is worthy mentioning the stable states are not only dependent on the length of trajectories, $t$, they are also dependent on the applied simulator itself. In other words, the stable states are dependent on the applied propagator in the simulations. It might provides a way in detecting the possible effects of thermostats in canonical-ensemble MD simulations. 

For any (small) $t$, ones can start from conformations sampled in a model, such as $U(r)$, to independently generate $t-$length MD trajectories in $V(r)$. The trajectory from an initial conformation, such as $r_{0}^{k}$, contributes to equilibrium properties of $V(r)$ with the weight $w_{k} = {\cal W}_{U,V}(r_{0}^{k})$. Here $U(r)$ is a cheaper and smoother approximation of $V(r)$ with the same resolution, $r$. For example, $U(r)$ is an all-atomic force field with a higher temperature while $V(r)$ is the {\it ab~initio} interaction with a lower temperature. 
It is a rather different challenge while a CG model, $U(x)$, is applied as the starting point, since 
 initial conformations of MD trajectories must be constructed by subtly adding into the missing degrees of freedom, $y$, so that the distribution of the formed conformations is known, thus the weights of the trajectories can be estimated. 
The subtle construction dominantly determines the accuracies and efficiencies of the current CG-based enhanced sampling methods, such as the resolution exchange method~\cite{LiuSLV2008}. 
However, as our discussion above, while $t$ is not very short, the initial conformations, $\{r_{0}^{k}\}$, are grouped into stable regions, the weights of all the trajectories started from the same region can be set as a constant, such as, $w_{k} = w[\alpha]$ if $r_{0}^{k} \in \alpha$. Here $\alpha$ indexes the stable regions. 
 It is an important improvement to replace the weights ${\cal W}_{U,V}(r_{0}^{k})$ with $w[\alpha]$, since the latter does not change while the initial conformations are shortly relaxed. 
 Therefore, we can arbitrarily add the missing degrees of freedom, $y$, into CG conformations sampled in $U(x)$ to form some $\{r^{k}\}$, then we (minimize $V(r)$ a few steps by constraining the CG variable, $x$, if it is necessary, and) run short (in comparison with $t$) normal MD simulations in $V(r)$ to relax these $\{r^{k}\}$ to $\{r_{n}^{k} \}$ to remove the possibly interatomic overlap. Finally, we independently generate multiple $t-$length trajectories in $V(r)$ started from $\{r_{n}^{k}\}$.  
Although the short relaxations makes the distribution of the initial conformations be unknown, it does not change the stable regions which the  conformations belong to. 
Here it is possible that conformations with same $x$ but different $y$ belong to different stable regions, thus trajectories from these conformations may have different weights. 
We can analyze and group $\{r_{n}^{k}\}$ into some regions, and estimate ensemble means as, 
\begin{eqnarray}
\langle A \rangle = \frac{\sum_{\alpha} w[\alpha] \sum_{r_{n}^{k} \in \alpha} \overline{A}[r_{n}^{k};t]}{\sum_{\alpha} n_{\alpha} w[\alpha] }, 
\label{mean-regions}
\end{eqnarray}
where $n_{\alpha}$ is the number of these $r_{n}^{k}$ inside the $\alpha^{th}$ region, 
and $\overline{A}[r_{n}^{k};t]$ is the mean of any $A(r)$ in the trajectory(-ies) started from $r_{n}^{k}$. 
Here $w[\alpha] \propto n_{\alpha}^{-1} \int_{\alpha} e^{-V(r)} dr$. 
A single trajectory (or multiple trajectories for getting better statistics) from an initial conformation, such as $r_{n}^{k}$, is applied to represent the sub-ensemble of trajectories, $[r_{n}^{k};t]$. 
 It is possible to estimate $w[\alpha]$ from the average of ${\cal W}_{U,V}(r)$ in the $\alpha$ region if $n_{\alpha}$ is sufficient large. 
However, we can more efficiently estimate $w[\alpha]$ by supposing each $t-$length MD trajectory reaches the local equilibrium in the stable region whose initial conformation belongs to. Although a small fraction of trajectories might transition out of their initial stable regions, the part of such a trajectory before the transition is still supposed to be long enough to reach the local equilibrium. 
Thus, instead of identifying stable regions of these initial conformations, the weights of trajectories (or more exactly, of sub-ensembles), $\{w_{i}\}$, can be directly estimated from eq.(\ref{trajectoryweight}). If a trajectory happens a transition, only the first part before the transition is applied in the estimate by using a positive $\Delta_{i}$ in eq.(\ref{trajectoryweight}). Here multiple trajectories from an initial conformation can be generated to get better estimate of the weight of the sub-ensemble, if it is necessary. 
Ensemble mean of any $A(r)$ is estimated as 
$\langle A \rangle = \frac{\sum_{k} w_{k} \overline{A}[r_{n}^{k};t]}{\sum_{i} w_{i}}$.  
 
 The initial conformations of trajectories can be grouped into clusters by calculating the directed distances among the trajectories. 
Each cluster, wherein the distances are smaller than a chosen threshold value, corresponds a stable region of the system in the time $t$ and in the applied simulator. 
Some of these trajectories might be supposed to happen transitions, for example, if they large deviate from the other states, or if a positive $\Delta$ is judged to apply in the calculation of $w_{i}$ from eq.(\ref{trajectoryweight}).   
 We can calculate the distance of the two ending parts of these trajectories from the other states to detect the corresponding transitions.   Generally, the deviation of a single conformation, $r^{k}$,   
from a sample $X$, is  
\begin{eqnarray}
s^{2}_{X,r^{k}} = g_{\mu\nu}(X)~\delta_{X}A^{\mu}(r^{k})~\delta_{X}A^{\nu}(r^{k}). 
\label{conformation-distance}
\end{eqnarray}
Here $\delta_{X}A^{\mu}(r^{k})$ is 
 the difference of the value of $A^{\mu}(r)$ at $r^{k}$ from its mean in the sample $X$. 
Then the deviation of multiple independent conformations, $\{r^{k}\}, k=1, \cdots, m$, from $X$ is, 
$s^{2}_{X,\{r^{k}\}} = \frac{1}{m^{2}} \sum_{k} s^{2}_{X,r^{k}}$.
Due to the finite sizes of samples, $s^{2}_{X,\{r^{k}\}}$ is in the order of $s^{2}_{c} =n (\frac{1}{m} + \frac{1}{M})$ rather than zero, even while the distribution of $\{r^{k}\}$ is same as that of $X$. Here $n$ is the number of the applied basis functions, and $M$ is the size of $X$. Thus, $s^{2}_{c}$ provides a reference in the cluster analyses of initial conformations. 
Therefore, without requiring a priori knowledge of stable states, the cluster analyses (CA) forms a network of stable states in original systems. 
The free energies of the stable states can be estimated from the weights of trajectories which started from the states, without requiring to know conformational boundaries of the states. 
 
In estimating weights of trajectories (or of sub-ensembles) from eq(\ref{trajectoryweight}), the relaxation of initial conformations should be short in comparison with the  length of each trajectory, $t$, so that the relaxation does not change the stable regions of these initial conformations. 
It is possible to generate trajectories from arbitrary initial conformations and estimate their weight. 
Consider an arbitrary conformational sample, $X=\{r_{a}^{k}\}$, we generate trajectories from $\{r_{a}^{k}\}$,  and group the conformations into stable regions. 
Although the distribution of $r_{a}^{k}$ is unknown, the weight function ${\cal W}_{X,V}(r)$ can be defined and be expanded based on eq.(\ref{omega-expansion}) with some unknown variables, $w[\alpha], \alpha = 1, 2, \cdots$. In the other hand, $w[\alpha]$ can be written as the average of ${\cal W}_{X,V}(r_{a}^{k})$ in the ${\alpha}^{th}$ region, thus 
\begin{eqnarray}
w[\alpha] = 1 + \frac{\sum_{\beta} {\Gamma}_{\alpha\beta} w[\beta]}{\sum_{\beta} n_{\beta} w[\beta]},
\label{weightequation}
\end{eqnarray}
where ${\Gamma}_{\alpha\beta} = g_{\mu\nu}(X)~\overline{ A^{\mu}(r_{a}^{k}) }[\alpha] ~ \sum_{r_{a}^{k} \in \beta} \overline{A^{\nu}}[r_{a}^{k};t]$. 
$\overline{A^{\mu}(r_{a}^{k})}[\alpha] = \frac{1}{n_{\alpha}} \sum_{r_{a}^{k} \in \alpha} A^{\mu}(r_{a}^{k})$, and $n_{\alpha}$ is the number of the initial conformations in the $\alpha$ region. 
Here the mean of $A^{\mu}$ in $X$ was already set as zero.   
While the size of $X$ is large, and $n_{\alpha}$ in each $\alpha$ region is large, $w[\alpha]$ can be estimated well from eq.(\ref{weightequation}), no matter how these initial conformations are constructed. 

 
The CG-WED-CA approach provides a scheme in analyzing stable conformational regions and ensemble dynamics within the total simulation time scale, as well as in enhanced sampling in complex systems, such as biological macromolecules: (i) construct CG models  and generate complete CG samples; (ii) construct initial conformations in original systems to start WED simulations and calculate weights of the MD trajectories; 
(iii) analyze stable states of the initial conformations ( and may compare with 
available experimental results, such as, native folded states and typical partially folded states in proteins);
(iv) statistically detect transitions among the states from the trajectories. 
In enhanced sampling, the scheme is much flexible and efficient in comparison with the previous methods, such as replica exchange method~\cite{EarlD2005} or its development, the resolution exchange method~\cite{LiuSLV2008}, where either many replicas are needed or the extra degrees of freedom must be subtly added so that the formed conformations satisfy a known distribution.  
The scheme also provides a start point to detect dynamics in longer time scale than the total simulation time, based on the slow dynamics techniques between two known ends, such as the transition path sampling~\cite{BolhuisCDG2002}. 
Ones also might find a controllable way to modify the probability of generating trajectories in the trajectory spaces so that the slow transition trajectories are more focused on. 

\begin{acknowledgments}
X. Z acknowledges the Max Planck Society(MPG) and the Korea Ministry of Education, Science and Technology(MEST) for the support of the Independent Junior Research Group at the Asia Pacific Center for Theoretical Physics (APCTP). 
He is grateful to Y. Jiang for stimulating discussions. 
\end{acknowledgments}




\begin{thebibliography}{13}

\bibitem{Elber2005} Ron~Elber, Curr. Opin. Struct. Biol. {\bf 15}, 151 (2005). 

\bibitem{VoterMG2002} A.~F.~Voter, F.~Montalenti, and T.~C.~Germann, Annu. Rev. Mater. Res. {\bf 32}, 321 (2002).

\bibitem{BolhuisCDG2002} P.~G.~Bolhuis, D.~Chandler, C.~Dellago, P.~L.~Geissler, Annu Rev. Phys. Chem. {\bf 53}, 291 (2002). 

\bibitem{SherwoodBS2008} P.~Cherwood, B.R. Brooks, and M. S. Sansom, Curr. Opin. Struct. Biol. {\bf 18}, 630 (2008).


\bibitem{MuellerPlathe2002} F.~Mueller~Plathe, Chem. Phys. Chem. {\bf 3}, 754 (2002).



\bibitem{KremerGroup} L.~Delle~Site, C.~F.~Abrams, A.~Alavi and K.~Kremer, Phys. Rev. Lett. {\bf 89}, 156103 (2002); X.~Zhou, D.~Andrienko,~L.~Delle~Site, and K.~Kremer, Europhys. Lett. {\bf 70}, 264 (2005); X.~Zhou, D.~Andrienko,~L.~Delle~Site, and K.~Kremer, J. Chem. Phys. {\bf 123}, 104904 (2005).


\bibitem{IzvekovV2005} S.~Izvekov, and G.~A.~Voth, J. Phys. Chem. B {\bf 109}, 2469 (2005);  
W.~G.~Noid {\it et al.} J. Chem. Phys. {\bf 128} 244114 (2008).


\bibitem{ZhouJZR2008} X. Zhou, Y. Jiang, H. Ziock, and S. Rasmussen, J. Chem. Phys. {\bf 128} 174107 (2008).


\bibitem{Voter1998} A.~F.~Voter, Phys. Rev. B {\bf 57}, R13985 (1998).  

\bibitem{ShirtsP2001} M.~R.~Shirts, and V.~S.~Pande, Phys. Rev. Lett. {\bf 86}, 4983 (2001). 

\bibitem{foldingathome} X.~Huang, G.~R.~Bowman, and V.~S.~Pande, J. Chem. Phys. {\bf 128} 205106 (2008); more related works can be found  at the webpage of folding@home.  



\bibitem{EarlD2005} D.~J.~Earl and M.~W.~Deem, Phys. Chem. Chem. Phys. {\bf 7}, 3910 (2005). 


 


\bibitem{LiuSLV2008} P.~Liu, Q.~Shi, E.~Lymn, and G.~A.~Voth, J. Chem. Phys. {\bf 129}, 114103 (2008).


\end{thebibliography}


 


\end{document}